\documentclass[10pt,twocolumn,letterpaper]{article}

\usepackage{cvpr}
\usepackage{times}
\usepackage{epsfig}
\usepackage{graphicx}
\usepackage{amsmath}
\usepackage{amssymb}
\usepackage{pifont}
\usepackage{commath}
\usepackage{subfigure}
\usepackage{wrapfig}

\newcommand{\cmark}{\ding{51}}%
\newcommand{\xmark}{\ding{55}}%
\DeclareMathOperator*{\argmax}{arg\,max}

\usepackage[pagebackref=true,breaklinks=true,letterpaper=true,colorlinks,bookmarks=false]{hyperref}

\cvprfinalcopy 

\def\cvprPaperID{****} 

\ifcvprfinal\fi
\begin{document}

\title{Towards Sharper First-Order Adversary with Quantized Gradients}

\author{Zhuanghua Liu\\
University of Technology Sydney\\
Amazon AWS\\
{\tt\small liuzhuanghua1991@gmail.com}
\and
Ivor W. Tsang\\
University of Technology Sydney\\
{\tt\small ivor.tsang@uts.edu.au}
}

\maketitle

\begin{abstract}
Despite the huge success of  Deep Neural Networks (DNNs) in a wide 
spectrum of machine learning and data mining tasks, recent research shows that this powerful
tool is susceptible to maliciously crafted adversarial examples. Up until now, adversarial training
 has been  the most successful
defense against adversarial attacks. To increase adversarial robustness, a  
DNN can be trained with a combination of benign  and adversarial examples generated by 
first-order methods. However, in  state-of-the-art first-order attacks, adversarial 
examples with sign gradients retain the sign information of each gradient
component but discard the relative magnitude between components.
 In this work, we replace sign gradients with quantized gradients.
Gradient quantization not only preserves the sign information, but  also keeps the relative
magnitude between  components. Experiments show white-box first-order attacks with 
quantized gradients outperform their variants with sign gradients on multiple datasets.
 Notably, our
BLOB\_QG attack achieves an accuracy of $88.32\%$ on the secret MNIST model from the  MNIST Challenge
and it outperforms all other methods on the leaderboard of white-box attacks.
\end{abstract}

\section{Introduction}

\begin{figure}[b]
  \begin{center}
    \includegraphics[width=0.18\textwidth]{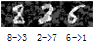}
  \end{center}
   \caption{Perturbed MNIST by PQGD}
   \label{fig:sample}
\end{figure}

Although Deep Neural Networks (DNN) are powerful tools in a wide range of applications, 
they are vulnerable to adversarial examples, which are crafted by adding small perturbations
to benign examples \cite{DBLP:journals/corr/SzegedyZSBEGF13}. Perturbations are so small
that   they are undetected by a human observer while they confuse well-trained
DNNs to make wrong predictions, see Figure \ref{fig:sample}. Consequently, the lack of robustness of DNNs  has 
raised significant concerns for safety in critical technologies such as autonomous driving and malware prevention
\cite{DBLP:conf/ccs/SharifBBR16, DBLP:journals/corr/EvtimovEFKLPRS17}.

One approach to increase adversarial robustness of DNNs is to train the model with data
augumented by adversarial examples. In particular, \cite{DBLP:conf/iclr/MadryMSTV18}
suggests the following Empirical Risk Minimization (ERM) as the objective:

\begin{equation}
\begin{aligned}
\min_{\theta} \quad & \rho(\theta)\\
\textrm{where} \quad  \rho(\theta)& = \mathbb{E}_{(x, y) \sim\mathcal{D}}\big[\max_{\delta \in \mathcal{S}}L(\theta, x+\delta,y)\big]\\
\end{aligned}
\label{eq:erm}
\end{equation}

Where $\theta$ denotes the parameters of DNNs, data examples $x$ with labels $y$ are from the data distribution
$\mathcal{D}$ and $L(\theta, x, y)$ is the loss function of DNN. $\mathcal{S}$ is a set of allowed perturbations
for crafting adversarial examples. In this paper, we consider the well studied $\ell_{\infty}$  perturbations, i.e.,
given some small threshold $\epsilon$, we have $\|\delta\|_{\infty} \leq \epsilon$ and it implies small visible changes
in images for each pixel.
Most of the first-order attacks used for  adversarial training
 are crafted under $\ell_{\infty}$  perturbations.
 Solving the inner maximization problem equals to finding  adversarial
examples that the neural network fails to classify correctly while the outer minimization problem tends to 
increase the robustness against such adversaries. 

\textit{Raw Gradients}\quad To solve the inner maximization problem of ERM in Eq. \ref{eq:erm}, one class of approach uses 
first-order methods to compute the perturbation $\delta$ and generate the adversarial example $x+ \delta$.
However, raw gradients $\nabla_x L(\theta, x,y)$ cannot be directly applied to images $x$, because 
magnitudes of raw gradients are at a different scale to the  image pixels. 
For example, for the MNIST dataset, most components of raw gradients
 concentrate at  values ranging from $10^{-5}$ to $10^{-4}$, while the pixel value of MNIST images 
 belongs to the set $\{0, 1/256, \ldots, 1\}$. 

\textit{Sign Gradients} \quad
To make use of  raw gradients and apply them to pixel values,
reseachers consider taking the sign
of each  raw gradient.
In particular, the  Fast Gradient Sign Method (FGSM) 
\cite{DBLP:journals/corr/GoodfellowSS14} linearizes the inner maximization problem with 

\begin{equation}
\label{eq:fgsm}
 x_{FGSM} = x + \epsilon \cdot \textrm{sgn} ( \nabla  L(\theta, x, y))
\end{equation}

Note that $\textrm{sgn} ( \nabla  L(\theta, x, y))$ takes the  sign of gradients
 $\nabla  L(\theta, x, y)$, i.e., gradient values belong to the set of $\{-1, 1\}$. 
 A more sophisticated approach to generate adversarial examples is to use multi-step FGSM
 with a smaller step size $\alpha$, e.g.,  Iterative FGSM  
\cite{DBLP:conf/iclr/KurakinGB17} and Projected Gradient Descent (PGD) attack 
\cite{DBLP:conf/iclr/MadryMSTV18}.
Evidence in \cite{DBLP:conf/iclr/MadryMSTV18} has shown that the PGD attack 
is the universal first-order adversary. 

\textit{Quantized Gradient} \quad 
One limitation of  sign gradients is that it discards the \textit{relative magnitude}
between each component of the raw gradient. 
When a component of raw gradient has its absolute value much larger than another component, 
it plays a more important role in solving the inner maximization problem. However,  
 sign gradients only make the two components share the same absolute value of 1.
Because a smaller step size $\alpha$ is used to generate adversarial examples,
we consider taking gradient values from a wider range of integers than $\{-1, 1\}$. 
Inspired by the success of  quantized gradients 
\cite{DBLP:conf/nips/AlistarhG0TV17} in distributed optimization, we introduce   Quantized
Gradients in first-order attacks to preserve the relative magnitude. Not only do quantized gradients retain the sign of 
each component,  they also round each component 
to an integer proportional to its magnitude 
 to preserve the relative magnitude. 
See Table \ref{tab:gradient_compare} for the comparison between raw gradients,
sign gradients and quantized gradients.

\begin{table*}[ht]
\begin{center}
\begin{tabular}{|c|c|c|c|}
\hline
\textbf{Gradient Scheme}& \textbf{Applicable on Pixels} & \textbf{Sign Preservation}&\textbf{Relative Magnitude Preservation} \\
\hline
Raw Gradients & \xmark & \cmark & \cmark  \\
Sign Gradients & \cmark & \cmark & \xmark  \\
Quantized Gradients & \cmark & \cmark & \cmark  \\
\hline
\end{tabular}
\caption{Comparison of  different gradient schemes}
\label{tab:gradient_compare}
\end{center}
\end{table*}

Our contributions are summarized as follows:

\begin{itemize}
  \item We introduce generalized quantized gradients which can replace the sign gradients in existing first order attacks.
  Specifically, in this paper, we propose Projected Quantized Gradients Descent (PQGD) and 
  DAA-BLOB with quantized gradients (BLOB\_QG)
  as extensions of state-of-the-art first-order attacks PGD and DAA-BLOB.
  \item We show that PQGD and BLOB\_QG are sharper attacks than PGD and DAA-BLOB, respectively, 
   on multiple datasets. In particular, BLOB\_QG achieves an accuracy of 88.32\% on 
   the secret MNIST model of the MNIST challenge and it outperforms all other methods   
   on the leaderboard of white-box attacks. See Section \ref{qg_res}.
  \item We study the effect of maximum allowed perturbation and number of steps on PQGD. Particularly, when 
  maximum allowed perturbation is larger or step number is smaller, 
  adversarial examples generated by PQGD are sharper than PGD. See details in Section \ref{qg_pert}
  and \ref{qg_step}.
  \item Our technique is simple to implement as it only requires tuning a single hyperparameter $b$.
\end{itemize}

\section{Preliminaries}

Given an $L-$class neural network classifier $f: \mathcal{R}^d \to \mathcal{R}^L$, for 
each $d-$dimensional data example $x$ and its corresponding ground truth label $y$, a
$f$ is trained to match the predicted label $\hat{y} = \argmax_{k} f_k(x)$ with the true label
$y$ where $f_k(x)$ is the $k$th component of $f(x)$. To attack a well-trained classifier,
an adversarial example $x_{adv}$ is crafted 
so that $\hat{y}_{adv} \neq \hat{y}$ where 
$\hat{y}_{adv} = \argmax_{k} f_k(x_{adv})$ given
 perturbations $\|x_{adv} - x\|_{\infty} \leq \epsilon$. 

Attacks can be divided into two main categories as  summarized below:

\begin{itemize}
\item{Targeted Attack and Untargeted Attack} \quad By specifying a label $y' \neq \hat{y}$, an adversarial
example $x_{adv}$ is generated with its predicted label $\hat{y}_{adv} = y'$ under the 
$\ell_{\infty}$ constraint. It is known as the targeted attack. On the other hand, an
untargeted attack only requires $\hat{y}_{adv}$ to be any label except $\hat{y}$.
\item{White-box Attack and Black-box Attack} \quad If the attack has access to all information of 
the network and training dataset, it is known as the white-box attack. FGSM and PGD attacks belong 
to this category because  gradients are computed from network parameters and dataset. In contrast,
the capability of attackers is more restricted in the black-box attack setting. In this scenario, attackers do not know
the network parameters and network architecture, also access to any large training dataset is forbidden.
See \cite{DBLP:journals/corr/PapernotMG16, DBLP:conf/ccs/PapernotMGJCS17} for a detailed description
of black-box attacks.
\end{itemize}

In this paper, we consider white-box untargeted attacks under $\ell_{\infty}$ perturbations.
We now present  a state-of-the-art first-order attack known as Projected Gradient Descent attack.

Projected Gradient Descent (PGD) \cite{DBLP:conf/iclr/MadryMSTV18}
 is shown to be the universal first-order adversary. That is, a neural 
network trained with adversarial examples crafted using PGD is robust to any first-order attack.
Given a chosen parameter $\alpha$, a training example $x$ and its label $y$,
PGD is similar to the Iterative FGSM attack 
\cite{DBLP:conf/iclr/KurakinGB17} and applies the following updates at  iteration $t$ : 
\begin{equation}
\label{eq:pgd}
x^{(t+1)} = \Pi_{clip}(x^{(t)} +  \alpha \cdot \textrm{sgn} ( \nabla  L(\theta, x^{(t)}, y)))
\end{equation}

Note that PGD is different from iterative FGSM in that $x_{0}$ is randomly perturbed in the vicinity
of $x$, while iterative FGSM has $x_0 = x$. On the other hand, FGSM uses $\epsilon$ as the step size 
while PGD uses $\alpha$ where $\alpha$ is much smaller than $\epsilon$ in practice. For example,
 to craft MNIST adversarial examples, one has $\epsilon = 0.3$ and $\alpha = 0.01$.

\section{Proposed Method}

In this section, we extend first-order attacks with quantized gradients.

\subsection{Motivation}

Because of the $\ell_{\infty}$ constraint on the perturbation, 
each component of the perturbation is valid to take any value from the interval of $[-\epsilon, \epsilon]$. 
Using the sign gradient allows the perturbation at each iteration to take  values from the set $\{-\alpha, +\alpha\}$. One challenge in PGD attacks
 is to determine other valid perturbations in each iteration  leads to better attack. 
 We seek for a solution to the problem. Note that  existing 
 works attempt to craft first-order attacks with sign gradients, so the absolute value of the perturbation at each iteration
 is equal to the step size $\alpha$. Our research uses different perturbation values for each component
  to craft a shaper attack, for this reason our work is orthogonal to existing research on first-order attacks.
 
  We first introduce the technique of gradient quantization and  propose a generalized gradient quantization
   to show that existing sign gradient attack can be replaced with the quantized gradient. We inspect the distribution
   of perturbation values of quantized gradients, sign gradients and raw gradients. Finally, we analyze the time complexity
   of our approach compared with vanilla sign gradients.

\subsection{Gradient Quantization}

Our approach attempts to preserve the relative magnitude between each component in the raw
gradient while it is still applicable on pixel value as shown in Table \ref{tab:gradient_compare}. 
Given a positive integer hyperparameter $b$, data example $x$ and its corresponding
label $y$, the loss function $L$ and network parameters $\theta$, 
we calculate quantized gradient $\mathbf{qg}$ as follows:

\begin{enumerate}
\label{qg_step}
\item Calculate the gradient of loss function $\mathbf{g} = \nabla_{x} L(\theta, x, y)$
\item Take the max value of components in the gradient $M = \max_i{\abs{g_i}}$
\item Return the quantized gradient  $\mathbf{qg} = \zeta(b \cdot g/M)$
\end{enumerate}.

Where function $\zeta(\cdot)$ aims to round each component to its nearest integer except
when the absolute value of the component is less than 1. Specifically, it takes the following form:

\begin{equation}
\zeta(v_i) = 
\begin{cases}
\textrm{sgn}(v_i)  & \text{if}\, \abs{v_i} < 1, \\
\textrm{round}({v_i}) & \text{otherwise.}
\end{cases}
\end{equation} 

The  $\zeta$ function enforces each component
of $\mathbf{qg}$ to be non-zero integer. Function $\textrm{round}({v_i})$ rounds $v_i$
to its nearest integer. 

\textbf{Relationship to Sign Gradients}\quad In the special case where $b=1$,
quantized gradient $\mathbf{qg}$ degenerates to sign gradient which is used
in the current FGSM and PGD framework.  

Note that  only one hyperparameter $b$ needs to be tuned when computing the quantized gradient
$\mathbf{qg}$.  When $b$ is set to a  large integer, the relative magnitude
between each component of the raw gradient is preserved. Each  component of the perturbation is 
rescaled to an integer belonging to $[-b, b]$.

Indeed, gradient quantization is equivalent to finding a scheme to assign different step size to each
component of the perturbation. We later show this method of assignment helps to  craft more effective adversarial examples, 
where fewer iterations are needed to generate sharper adversarial examples than sign gradients.
 For adversarial training, it is too expensive
to augument adversarial examples which needs many iterations to generate. If 
adversarial examples of similar effectiveness can be generated using fewer iterations, 
adversarial training can  benefit from such examples.

\textbf{Connection to QSGD}\quad We briefly review how quantized gradients are used in distributed optimization.
In distributed optimization, one of the   bottlenecks in performance is the limited bandwidth required to collect gradients from several
parallel machines. Quantized Stochastic Gradient Descent (QSGD)\cite{DBLP:conf/nips/AlistarhG0TV17}
is proposed to conserve the bandwidth with a good convergence
gurantee.  To be specific, each component of the gradient is quantized and randomly rounded to a discrete set of values.
Given any gradient $\mathbf{v} \in \mathcal{R}^d$ and a hyperparameter $s$, the quantized gradient $Q_s(\mathbf{v})$
is defined as:
\begin{equation}
Q_s(v_i) = \|\mathbf{v}\|_2 \cdot sgn(v_i) \cdot \xi_i(\mathbf{v}, s)
\end{equation}

Given some positive  integer $l$ such that $\abs{v_i}/ \|\mathbf{v}\|_2 \in \big[ l/s, (l+1)/s\big]$, $\xi_i(\mathbf{v}, s)$
is given by

\begin{equation}
\xi_i(\mathbf{v}, s) = 
\begin{cases}
l/s  & \text{with probability}\, p \\
(l+1)/s & \text{otherwise}
\end{cases}
\end{equation}

where $p = 1-(\frac{\abs{v_i}}{\|\mathbf{v}\|_2}\cdot s - l)$.
 Experiments show that QSGD significantly outperforms its full-precision variant in terms of convergence
speed.

Although quantized gradients used in attacks bear some resemblance to QSGD, 
there are a few differences between the two quantization techniques.

\begin{itemize}
\item QSGD is a randomized approach whereas our method is deterministic. In particular, QSGD randomly 
rounds every component to the nearest floor integer or  ceiling integer, while our approach rounds every 
component using the deterministic function $\zeta(\cdot)$. 
\item QSGD uses $\ell2$-norm of the gradient to divide each component. On the other hand, we 
enforce the $\ell_{\infty}$ constraint
on the perturbation. So we use $\|\mathbf{g}\|_{\infty} = \max_i{\abs{g_i}}$ to divide each component.
\item QSGD has its component $Q_s(v_i) \in [0, 1]$, while our approach constrains each component of
 $\mathbf{qg}$  as any non-zero integer. 
\end{itemize}

\subsection{Generalized Gradient Quantization}

Here, we introduce generalized gradient quantization for first-order attacks.

Recall that at each iteration PGD update has the form of Eq. \ref{eq:pgd}. To
integrate quantized gradients with PGD, we  replace the sign gradient 
$\text{sgn}(\nabla_{x} L(\theta, x, y))$ with $\mathbf{qg}$. It takes the 
following form:

\begin{equation}
\label{eq:pqgd}
x^{(t+1)} = \Pi_{clip}(x^{(t)} +  \alpha \cdot \mathbf{qg})
\end{equation}

We name our approach Projected Quantized Gradient Descent (PQGD).
To facilitate other existing first-order attacks with sign gradients, 
i.e., for attacks with the form    
$x^{(t+1)} = \Pi_{clip}(x^{(t)} +  \alpha \cdot sgn(\cdot))$,
we replace  $sgn(\cdot)$ with
the quantized gradients to improve the effectiveness of attacks.
We call this approach generalized gradient quantization.

 Distributionally Adversarial Attack (DAA) is an example of
a state-of-the-art first-order white-box attack, which appears on
the leaderboard of the MNIST challenge.
While DAA is proposed to solve the optimal adversarial data
distribution \cite{DBLP:journals/corr/abs-1808-05537},
PGD views each data example independently. 
 DAA is interpreted as Wasserstein Gradient Flows, and the update
of DAA using the Lagranigian Blob Method (DAA-BLOB) at iteration $t$ is given as:

\begin{equation}
\begin{aligned}
\mathbf{eg} = \nabla_{ x^{(t)}_i}  &L(\theta, x^{(t)}_i, y_i) + \\ 
\frac{c}{M} \big[ \sum_{j=1}^{M} &K(x^{(t)}_i, x^{(t)}_j ) \nabla_{ x^{(t)}_j}  L(\theta, x^{(t)}_j, y_j)+ \\
\nabla_{ x^{(t)}_j}& K( x^{(t)}_i, x^{(t)}_j) \big]
\end{aligned}
\end{equation}

\begin{equation}
\label{eq:daa}
\begin{aligned}
x^{(t+1)}_i = \Pi_{clip}(x^{(t)}_i +  \alpha \cdot \textrm{sgn}(\mathbf{eg})
\end{aligned}
\end{equation}

where $K(\cdot, \cdot)$ is a kernel function.
We define the following quantized gradient for DAA-BLOB.

\begin{equation}
\mathbf{eqg} =  \zeta(b \cdot \mathbf{eg}/M_{eg}) \quad \text{where} \quad M_{eg} = \max_i{\abs{eg_i}}.
\end{equation}
 DAA-BLOB with quantized gradient (BLOB\_QG) has the following update in each iteration:

\begin{equation}
\label{eq:blob_qg}
x^{(t+1)} = \Pi_{clip}(x^{(t)} +  \alpha \cdot \mathbf{eqg})
\end{equation}

We name this approach DAA-BLOB with quantized gradient (BLOB\_QG).

\subsection{Inspection of Quantized Gradient}

\begin{figure*}[ht]
\subfigure[sign gradients]{\includegraphics[width=0.25\textwidth]{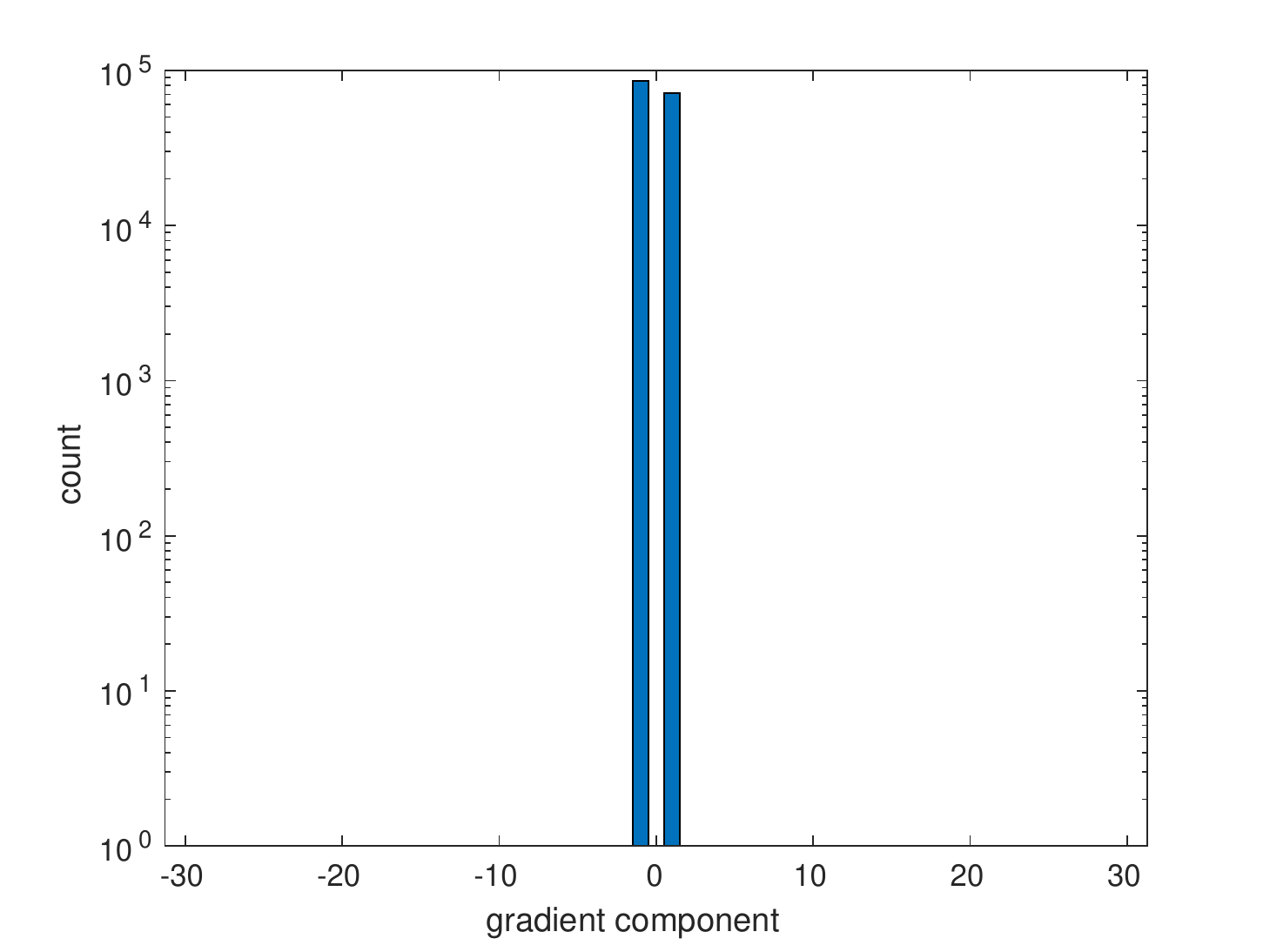} \label{fig:grad_1}}
\subfigure[quantized gradients (b=100)]{\includegraphics[width=0.25\textwidth]{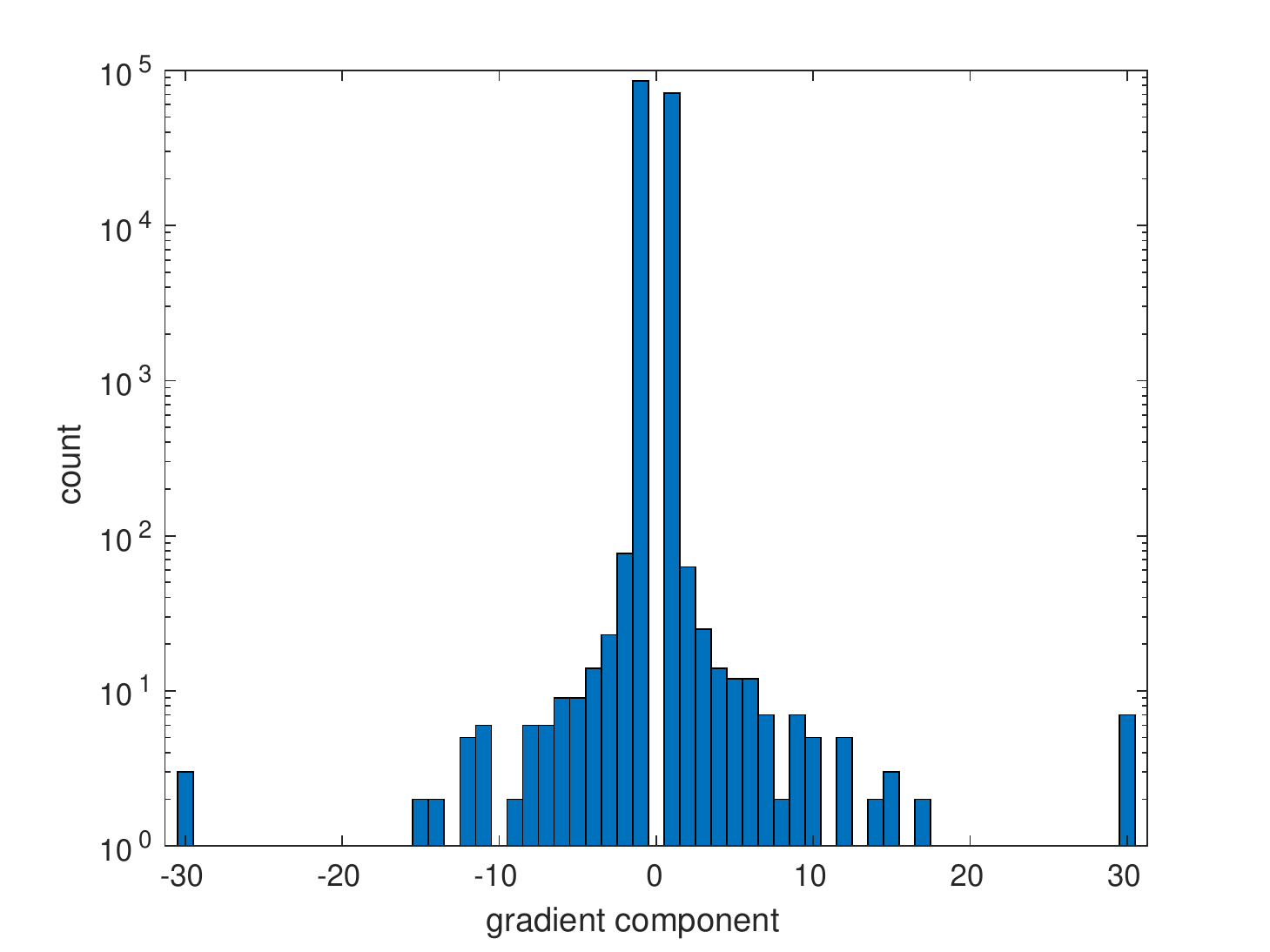} \label{fig:grad_100}}
\subfigure[quantized gradients (b=200)]{\includegraphics[width=0.25\textwidth]{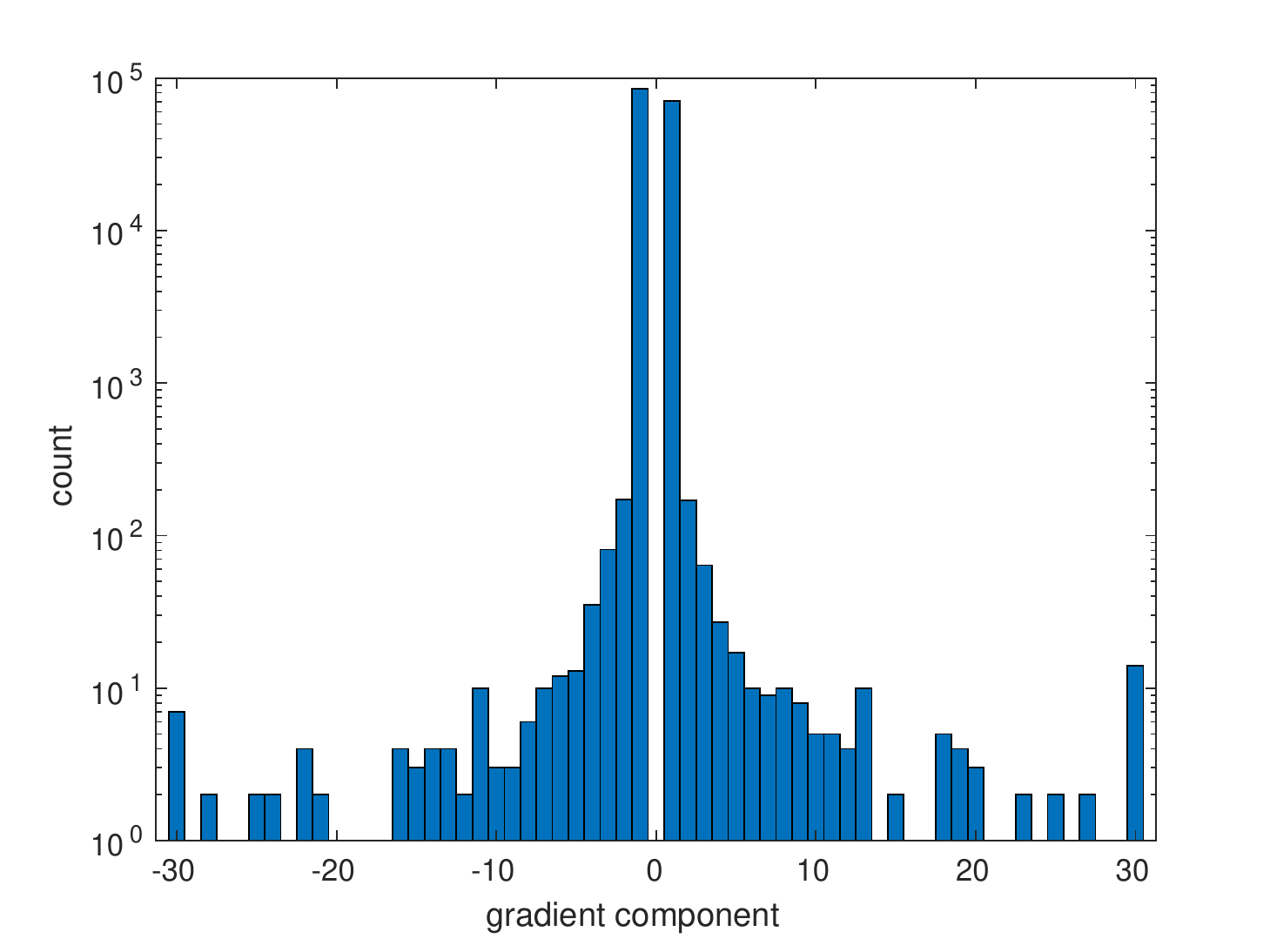} \label{fig:grad_200}}
\subfigure[quantized gradients (b=500)]{\includegraphics[width=0.25\textwidth]{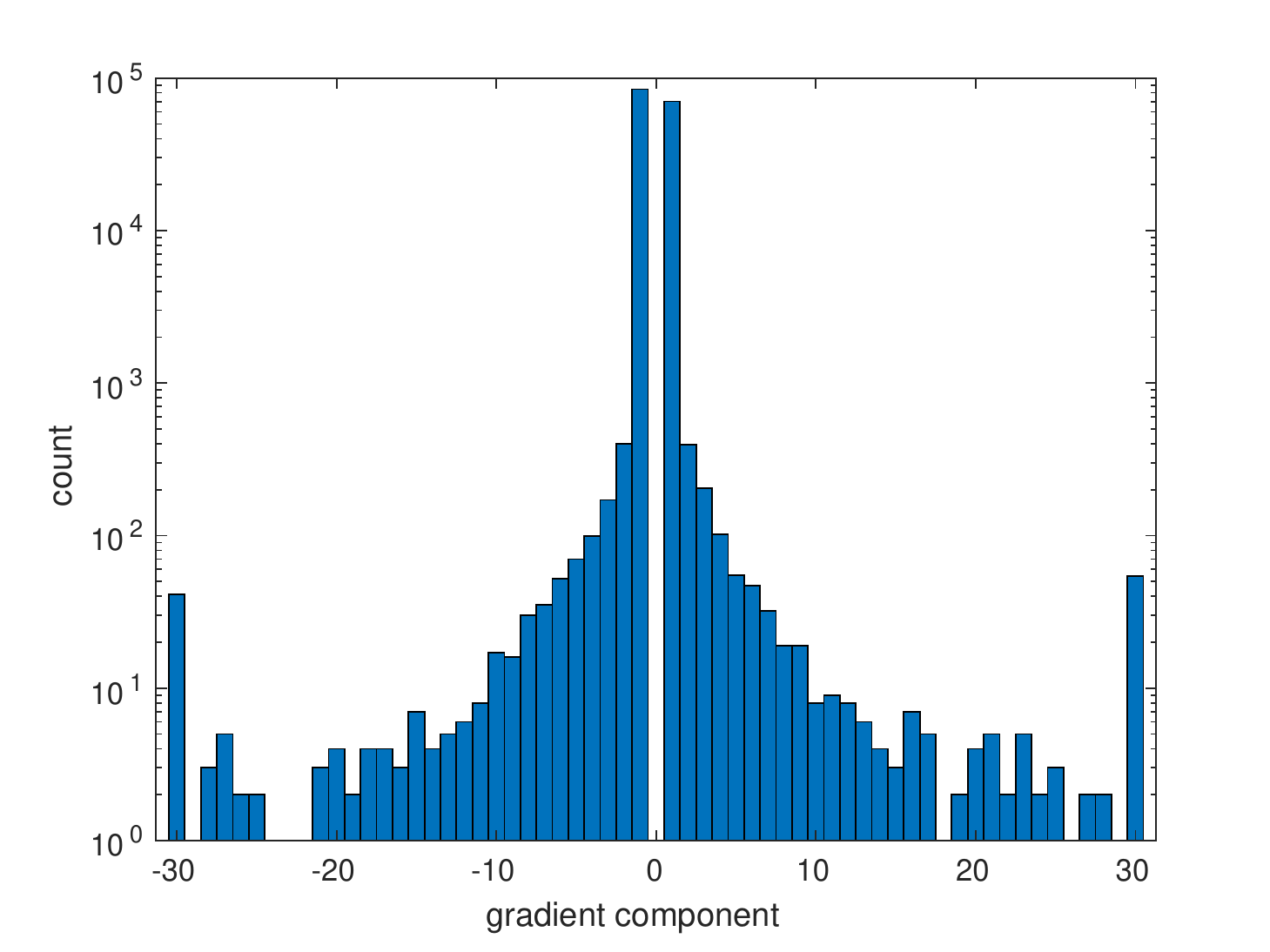} \label{fig:grad_500}}
\hfill
\subfigure[quantized gradients (b=1000)]{\includegraphics[width=0.25\textwidth]{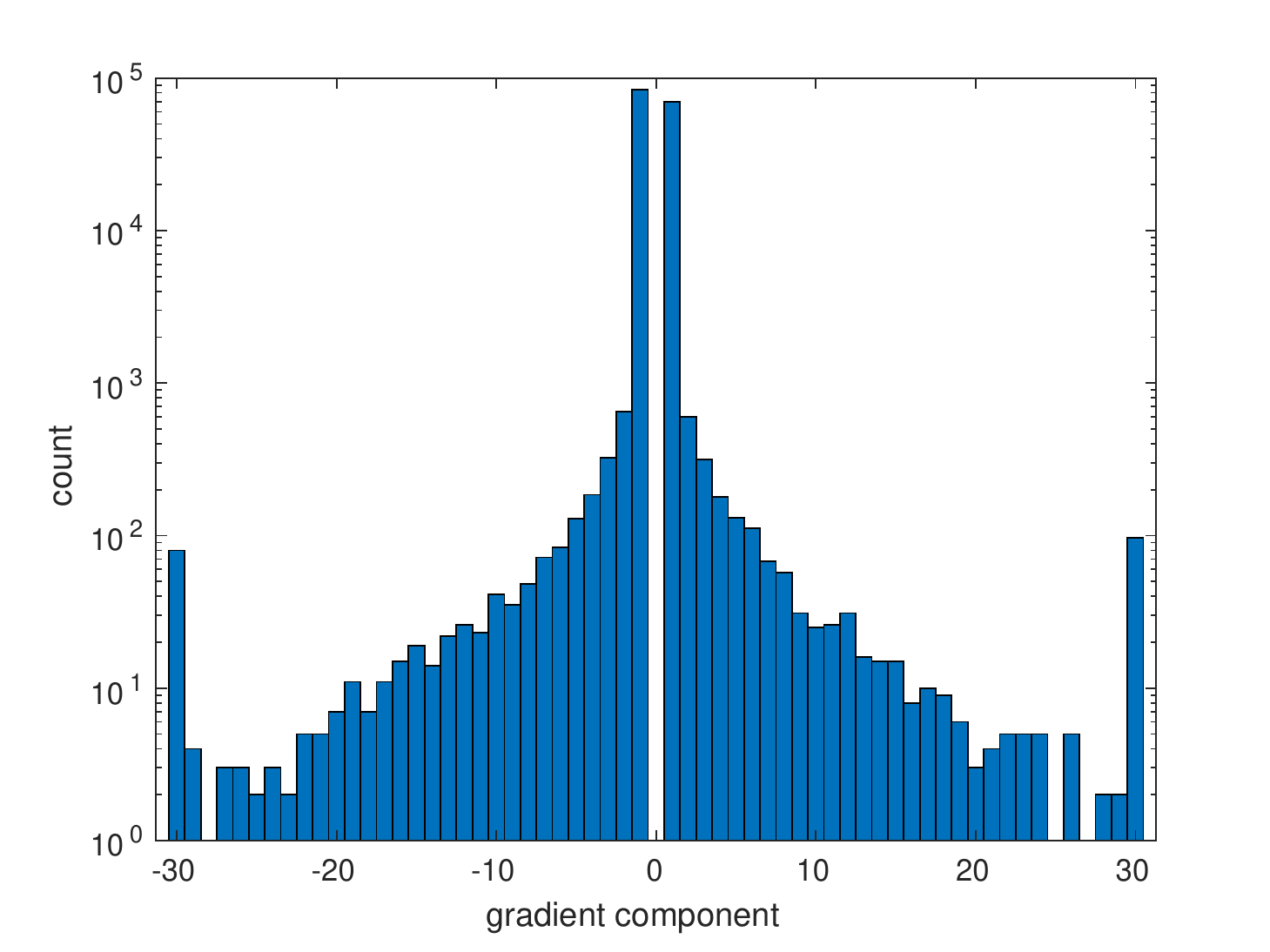} \label{fig:grad_1000}}
\hfill
\subfigure[raw gradient]{\includegraphics[width=0.25\textwidth]{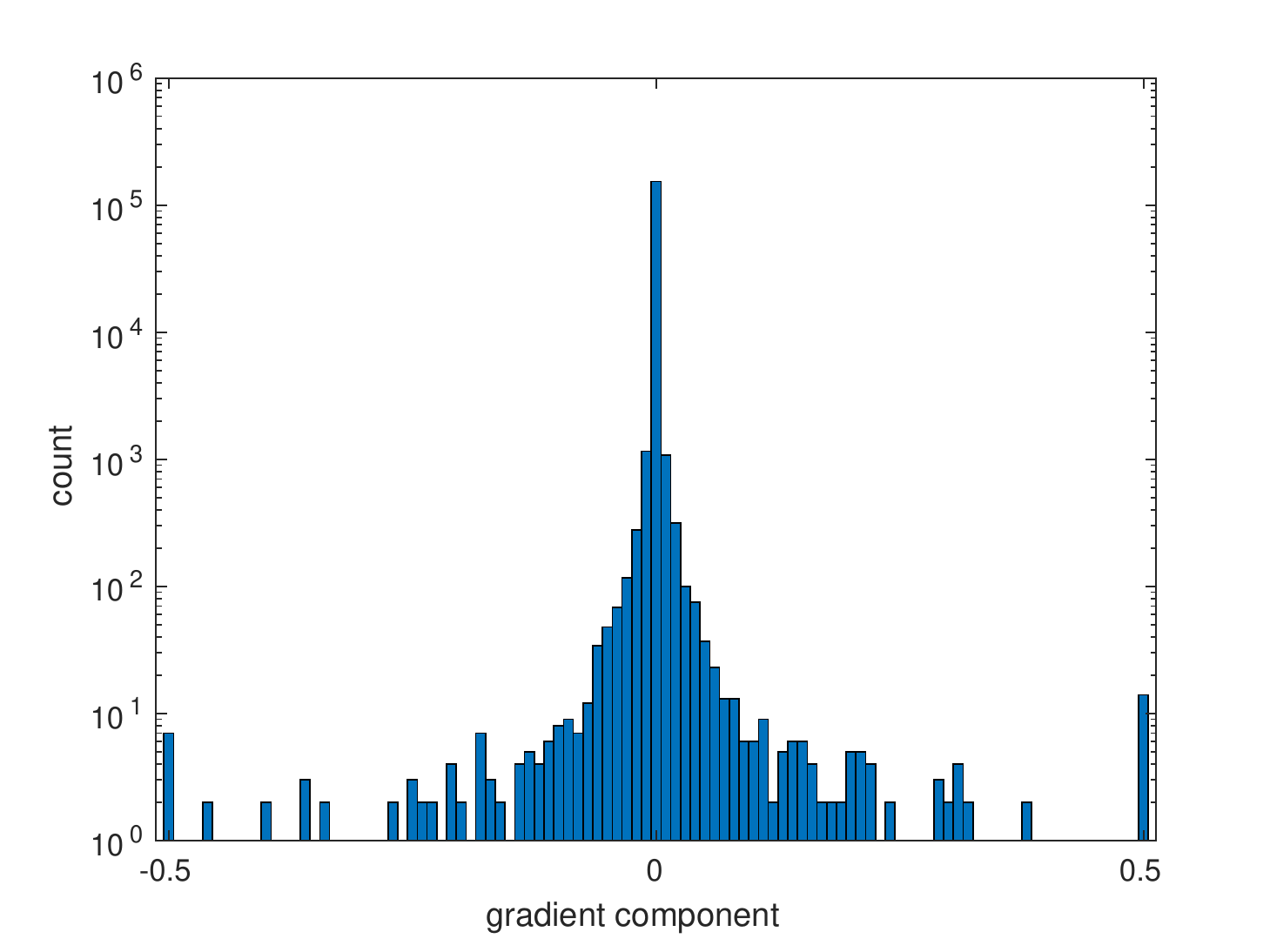} \label{fig:raw_gradient}}
\caption{Histograms of sign gradients, raw gradient and  quantized gradients of MNIST model with various values of $b$.}
\label{fig:grad_ins}
\end{figure*}

We plot the histogram of sign gradients,
raw gradients and quantized gradients of the adversarial trained
 MNIST model with various  values of $b$ (Fig. \ref{fig:grad_ins}). Recall that an iterative attack has a projection operation at the end
of each iteration. Therefore, the value change of each gradient component should be within the range of 
$[-\epsilon/\alpha, \epsilon/\alpha]$. Using MNIST as an example, we have a maximum allowed perturbation of $\epsilon=0.3$ and 
the learning rate of $\alpha=0.01$, so for the absolute value of  gradient component larger than 
$\frac{\epsilon}{\alpha}=30$, the value will be clipped inside the range of $[-30, 30]$.

Figure \ref{fig:grad_1} shows the histogram of the sign gradient, which is  used in PGD and 
other first-order attacks. Values of components of sign gradients concentrate on $\{-1, 1\}$ 
because only the sign of each component is preserved. In Figure \ref{fig:grad_100}
to Figure \ref{fig:grad_1000}, quantized gradients  are  spread out on the range of $[-30, 30]$ and have 
a  shape similar to the raw gradients shown in Figure \ref{fig:raw_gradient}.
Not only is
the sign of each component  kept for quantized gradients, 
but the relative magnitude between each component is preserved. 

Recall that we use function $\zeta(\cdot)$  to constrain each gradient component as non-zero integers. Thus,
if we do not apply the constraint, most of the gradient components will have a value of $0$. Accordingly these 
components do not  contribute to the  crafting of  adversarial examples. Therefore we enforce the components
to be non-zero integers and they are restricted to the range $\{-1, 1\}$, whose count dominates count of the other
integer values within the range $[-30, 30]$ as shown in Figure \ref{fig:grad_ins}.

\subsection{Complexity Analysis}

We analyze the computational complexity of the quantized gradient here. Compared with vanilla
 gradients $\nabla_x L(\theta, x, y)$ has a time complexity of
at least $\mathcal{O}(d)$ where $d$ is the dimension of data point $x$. Since computing the maximum $M$ and applying the function
$\zeta$ takes time of $\mathcal{O}(d)$. Because of this, computing the quantized gradient shares
the same time complexity as computing the raw gradient. Therefore, our technique
introduces low computational overheads.

In addition, our approach of gradient quantization is easy to implement. For instance, in our python implementation, 
we use around 10 lines of code.

\section{Experiment}

The purpose of our experiment is to test the effectiveness of existing first-order attacks with 
quantized gradient (PQGD and BLOB\_QG) on adversarial trained models compared to
 the attacks with sign gradients, e.g., PGD and DAA-BLOB.

\subsection{Experiment Settings}

 Our code is based on
the implementation of PGD and DAA-BLOB. The code of PGD can be found at
\href{https://github.com/MadryLab/mnist\_challenge}{code site}
and the implementation of DAA-BLOB can be found at 
\href{https://github.com/tianzheng4/Distributionally-Adversarial-Attack}{code site}.
 We use  PGD, DAA-BLOB, Interval Attacks 
\cite{DBLP:journals/corr/abs-1811-02625} and Surrogate attack \cite{gowal2019alternative} as the baselines
in our experiment.

We compare our methods and baseline on the following datasets: MNIST
\cite{lecun1998gradient}, Cifar10\cite{krizhevsky2009learning},  and 
Fashion-MNIST\cite{xiao2017fashion}. We mirror the architecture for MNIST and
Cifar10 from the architecture used in \cite{DBLP:conf/iclr/MadryMSTV18}
 while we use the same architecture
for Fashion-MNIST as that in \cite{DBLP:journals/corr/abs-1808-05537}. 
We use  default training configurations
including batch size, learning rate and optimization settings. 
Cross entropy loss is used in all of our experiments.

\textbf{Practical Issues} \quad Since PGD and DAA-BLOB use randomly perturbed
data examples to craft adversarial examples,  adversarial examples  generated in different runs are not exactly the same. 
An approach to take advantage of all different runs is to merge  various adversarial
examples that are misclassified by the classifier during different restarts.

\subsection{Empirical Results}\label{qg_res}

\begin{table*}[ht]
\begin{center}
\begin{tabular}{|c|c|c|c|c|c|c|c|c|c|c|c|}
\hline
\multicolumn{2}{|c|}{\textbf{PGD}}&\multicolumn{2}{|c|}{\textbf{PQGD\_100}}&\multicolumn{2}{|c|}{\textbf{PQGD\_200}}  &\multicolumn{2}{|c|}{\textbf{BLOB}} &\multicolumn{2}{|c|}{\textbf{BLOB\_QG\_100}} &\multicolumn{2}{|c|}{\textbf{BLOB\_QG\_200}}\\
\hline 
\textbf{Worst} & \textbf{Avg}& \textbf{Worst}& \textbf{Avg} & \textbf{Worst}& \textbf{Avg} & \textbf{Worst}& \textbf{Avg} & \textbf{Worst}& \textbf{Avg} & \textbf{Worst}& \textbf{Avg}\\
\hline
92.56\% & 92.79\% & 92.27\% & 92.37\% & 92.17\% & 92.33\%   & 90.48\% & 90.56\% &\textbf{90.07\%} & 90.18\%&90.08\% & \textbf{90.15\%}\\
\hline
\end{tabular}
\caption{ Performance Comparison of MIT MadryLab's secret MNIST model without restart under 200-step attacks
(0.3/1.0 $\ell_{\infty}$ perturbations). PQGD\_100 denotes PQGD with $b = 100$. BLOB is the DAA-BLOB
approach. BLOB\_QG\_100 denotes DAA-BLOB with quantized gradients and $b=100$. }
\label{tab:mnist_single}
\end{center}
\end{table*}

\begin{table*}[ht]
\begin{center}
\begin{tabular}{|c|c|c|c|c|c|c|}
\hline
\textbf{PGD}& \textbf{Interval Attack} & \textbf{BLOB} & \textbf{Surrogate}&\textbf{BLOB\_QG\_100}&\textbf{BLOB\_QG\_200} & \textbf{BLOB\_QG\_200 (100 runs)} \\
\hline
89.49\% & 88.42\% & 88.56\% & 88.36\% & 88.40\% &  88.35\%  &    \textbf{88.32\%}  \\
\hline
\end{tabular}
\caption{Empirical worst-case accuracy of MIT MadryLab's secret MNIST model  under 200-step attacks
(0.3/1.0 $\ell_{\infty}$ perturbations).  
In the last entry, we have BLOB\_QG with 100 restarts. All other approaches are with 50 restarts.  
BLOB is the DAA-BLOB
approach. BLOB\_QG\_100 denotes DAA-BLOB with quantized gradients and $b=100$. }
\label{tab:mnist_multi}
\end{center}
\end{table*}

\textbf{MNIST}\quad
We present the result of comparisons between our approach and baseline attacks on secret MNIST model in 
Tables \ref{tab:mnist_single} and \ref{tab:mnist_multi}. 
The secret MNIST model is taken from the Madry Lab's MNIST-Challenge competition. 
We test  attacks with quantized gradients with various values of $b$.

In Table \ref{tab:mnist_single}, we compute the worst accuracy and average accuracy of five independent runs
under a 200-step attack for each approach on the secret MNIST model.
 It shows  results consistent with experiments
from previous works \cite{DBLP:journals/corr/abs-1808-05537} where DAA-BLOB outperforms PGD.
Replacing sign gradients with quantized gradients  boosts the performance of PGD and DAA-BLOB. Specifically,  using quantized gradients in PGD decreases the worst accuracy of  the secret MNIST model from 
92.56\% to 92.17\% while using quantized gradients in DAA-BLOB decreases the worst accuracy from 
90.48\% to 90.07\%. 

We compare BLOB\_QG
 with state-of-the-art methods appearing on the leaderboard of the MNIST challenge. Our  BLOB\_QG approach
 with $b=200$ and 100 restarts achieves a worst accuracy of 
 $88.32\%$ and  outperforms all other methods on the leaderboard of white-box attacks (In Table \ref{tab:mnist_multi}).

\begin{table*}[ht]
\begin{center}
\begin{tabular}{|c|c|c|c|c|c|c|c|c|c|c|c|}
\hline
\multicolumn{2}{|c|}{\textbf{PGD}}&\multicolumn{2}{|c|}{\textbf{PQGD\_100}}&\multicolumn{2}{|c|}{\textbf{PQGD\_200}}  &\multicolumn{2}{|c|}{\textbf{BLOB}} &\multicolumn{2}{|c|}{\textbf{BLOB\_QG\_100}} &\multicolumn{2}{|c|}{\textbf{BLOB\_QG\_200}}\\
\hline 
\textbf{Worst} & \textbf{Avg}& \textbf{Worst}& \textbf{Avg} & \textbf{Worst}& \textbf{Avg} & \textbf{Worst}& \textbf{Avg} & \textbf{Worst}& \textbf{Avg} & \textbf{Worst}& \textbf{Avg}\\
\hline
71.11\% & 71.21\% & 70.79\% & 70.85\% & 70.66\% & 70.75\%   & 67.90\% & 67.98\% &67.33\% & 67.45\%& \textbf{67.30\%} & \textbf{67.39\%}\\
\hline
\end{tabular}
\caption{Performance Comparison of the adversarial trained Fashion-MNIST model without restart under 100-step attacks 
(0.2/1.0 $\ell_{\infty}$ perturbations). PQGD\_100 denotes PQGD with $b = 100$. BLOB is the DAA-BLOB
approach. BLOB\_QG\_100 denotes DAA-BLOB with quantized gradients and $b=100$. }
\label{tab:fmnist_single}
\end{center}
\end{table*}

\begin{table*}[ht]
\begin{center}
\begin{tabular}{|c|c|c|c|c|}
\hline
\textbf{BLOB}&\textbf{BLOB\_QG\_100}&\textbf{BLOB\_QG\_200} &\textbf{BLOB\_QG\_500} &\textbf{BLOB\_QG\_1000} \\
\hline
66.24\% & 65.78\% & 65.69\% & \textbf{65.61\%} &  65.64\%   \\
\hline
\end{tabular}
\caption{Empirical worst-case accuracy of the adversarial trained  Fashion-MNIST model  under 200-step attacks with 10 restarts
(0.2/1.0 $\ell_{\infty}$ perturbations).  BLOB is the DAA-BLOB
approach. BLOB\_QG\_100 denotes DAA-BLOB with quantized gradients and $b=100$. }
\label{tab:fmnist_multi}
\end{center}
\end{table*}

\textbf{Fashion-MNIST} We compare PQGD and BLOB-QG with baseline attacks 
on an adversarial trained Fashion-MNIST model in Tables 
\ref{tab:fmnist_single} and \ref{tab:fmnist_multi}. The adversarial Fashion-MNIST
model is trained with adversarial examples generated by a 40-step PGD attack. 

We observe  results of Fashion-MNIST, which are consistent with those of MNIST.
Existing attacks with quantized gradients create  
sharper attacks than those with sign gradients.

\begin{table*}[ht]
\begin{center}
\begin{tabular}{|c|c|c|c|c|c|}
\hline
\textbf{Perturbation} &\textbf{PGD}&\textbf{PQGD\_100}&\textbf{PQGD\_200} &\textbf{PQGD\_500} &\textbf{PQGD\_1000} \\
\hline
8.0/255.0 & \textbf{45.31\%} & 45.32\% & 45.32\% & 45.33\% &  45.57\%   \\
\hline
16.0/255.0 & 14.02\% & 14.24\% & 14.11\% & 13.98\% &  \textbf{13.87\%}   \\
\hline
\end{tabular}
\caption{Performance Comparison of the adversarial trained  Cifar10 model from Madry Lab with no restart under 100-step attack
(with 8.0 and 16.0/255.0 $\ell_{\infty} maximum allowed perturbations$).  PQGD\_100 denotes PQGD with $b=100$. }
\label{tab:cifar_res}
\end{center}
\end{table*}

\textbf{Cifar10} \quad
We compare the performance of adversarial examples generated by
 PGD and PQGD on the adversarial trained Cifar10 model from the
Madry Lab (in Table \ref{tab:cifar_res}). We test a single run of 100-step attacks of PGD and PQGD
 under 8.0/255.0 and 16.0/255.0 $\ell_{\infty}$ maximum allowed perturbations.

As shown in Table \ref{tab:cifar_res}, for a maximum allowed perturbation of 8.0/255.0. PQGD with $b=100, 200, 500$
 achieves a similar performance to PGD. However, PQGD with $b=1000$ produces a worse result than PGD with  an accuracy of $45.57\%$.
  On the other hand,  for maximum allowed perturbations of 16.0/255.0,  PQGD
decreases the accuracy of adversarial trained models from $14.02\%$ to $13.87\%$. Interestingly,
 PQGD with $b=1000$ outperforms PGD under the $\ell_{\infty}$ maximum allowed perturbation of 16.0/255.0 
whereas PGD outperforms PQGD under the maximum allowed perturbation of 8.0/255.0. 

\subsection{Effect Of Maximum Allowed Perturbation On Gradient Quantization}\label{qg_pert}

\begin{figure*}[ht]
\hspace{2.6cm}
\subfigure[MNIST]{\includegraphics[width=0.25\textwidth]{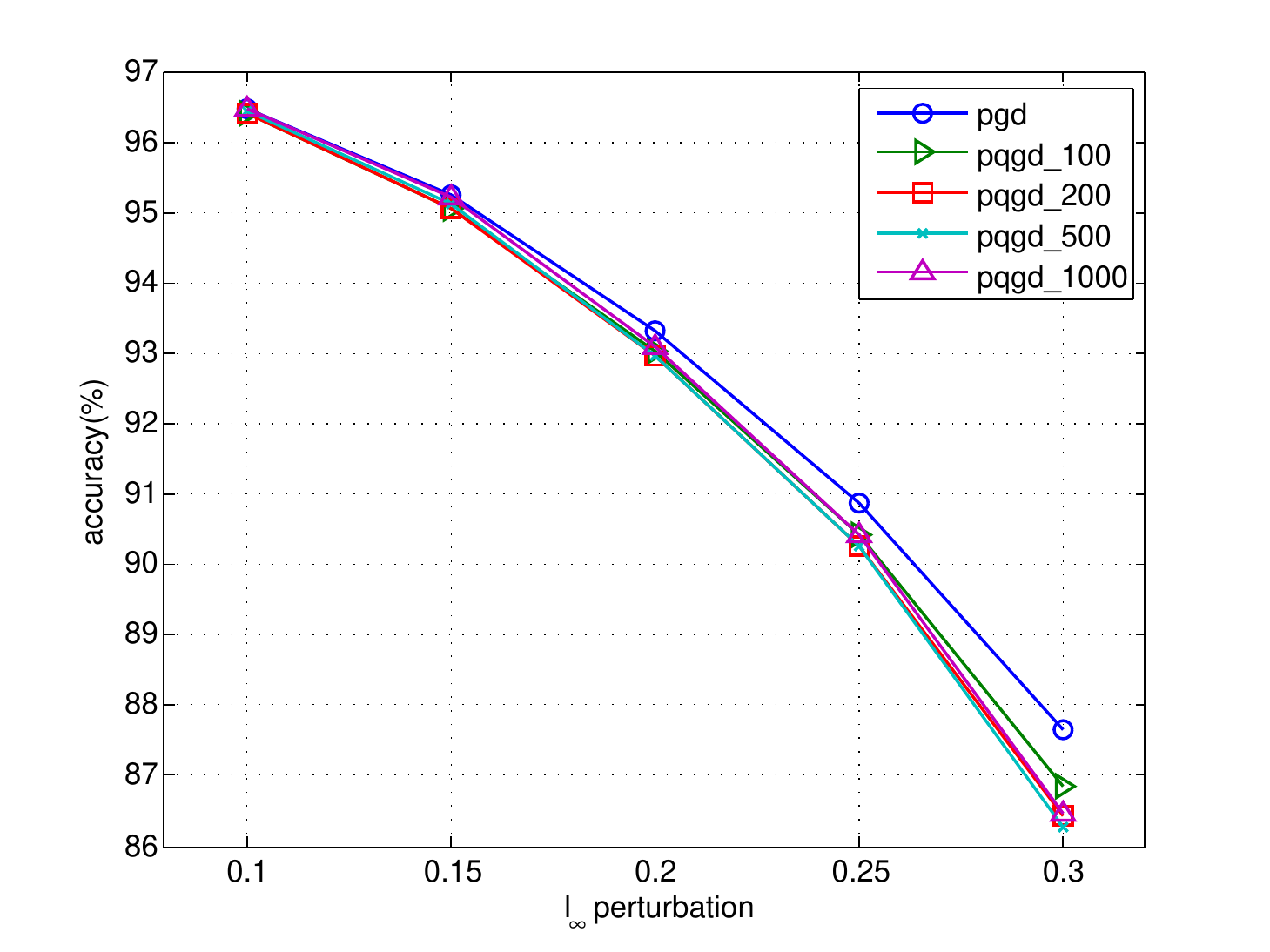} \label{fig:mnist_pert}}
\hspace{3cm}
\subfigure[Fashion-MNIST]{\includegraphics[width=0.25\textwidth]{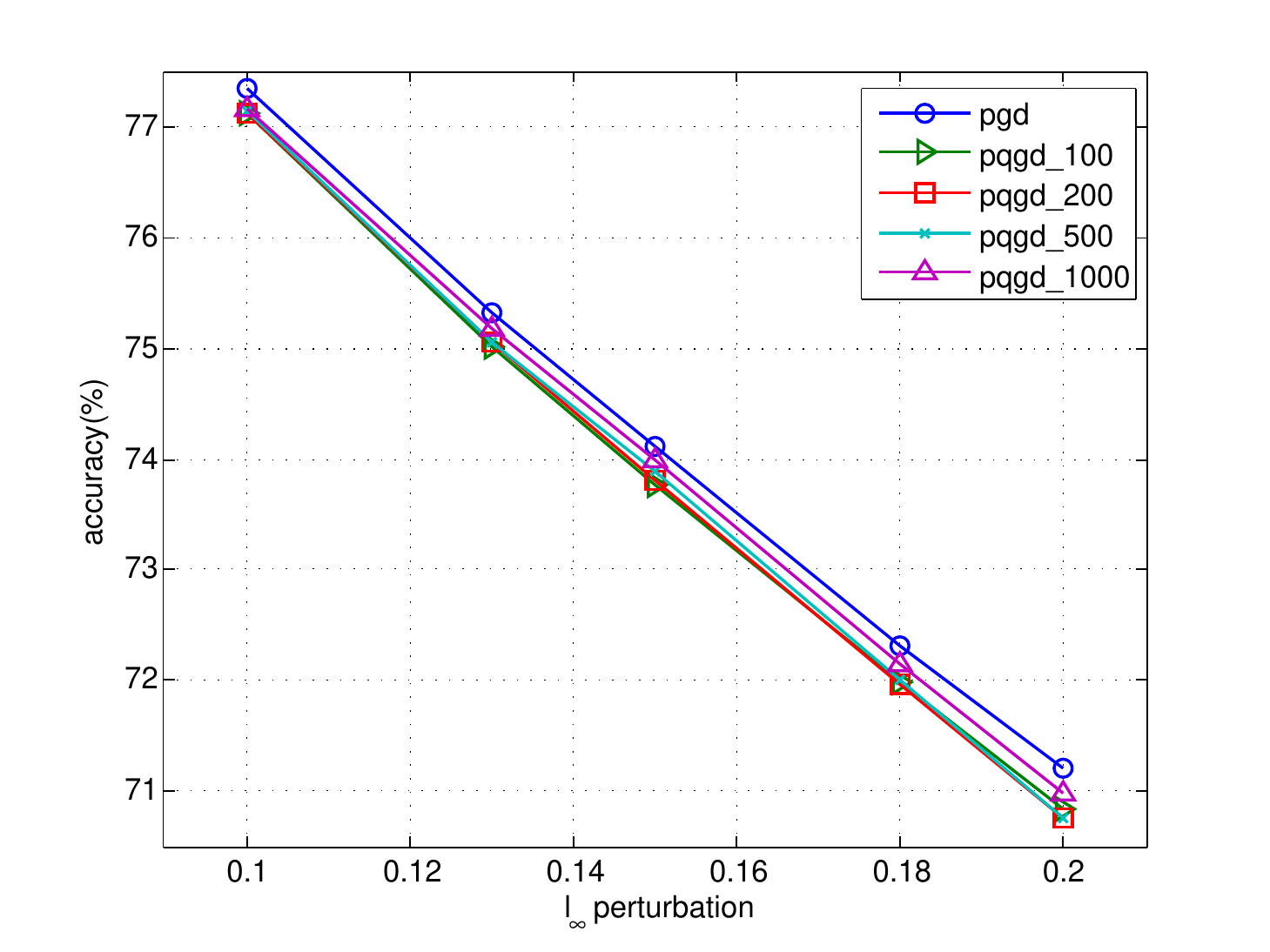} \label{fig:fmnist_pert}}
\hspace{2.6cm}
\caption{Effect of Maximum Allowed Perturbation on the averaged accuracy of  applying PQGD and PGD attacks on adversarial trained models}
\label{fig:pert}
\end{figure*}

We plot the average accuracy of five independent runs of 100-step PQGD and PGD with different
maximum allowed perturbations on the MNIST and Fashion-MNIST
dataset (Fig. \ref{fig:pert}).

The effect of maximum allowed perturbations on the performance of gradient quantization on the
MNIST dataset is shown in Figure \ref{fig:mnist_pert}. Here we test PGD and PQGD on an adversarial trained MNIST model.
The results show that when maximum allowed perturbation is small, i.e., 0.1/1.0 $\ell_{\infty}$ perturbation, 
PGD and PQGD with various $b$ achieve similar accuracy on the adversarial trained MNIST model.
When maximum allowed perturbation increases, PQGD with various $b$ is sharper than PGD. 
Consequently, when the maximum allowed perturbation is sufficiently large, i.e., 0.3/1.0 $\ell_{\infty}$ perturbation, PQGD
with several choices of $b$ outperforms PGD by a small margin.

The effect of  the maximum allowed perturbation on the performance of the gradient quantization
on the Fashion-MNIST dataset (Fig.\ref{fig:fmnist_pert}). For the Fashion-MNIST dataset, we obtain a  similar outcome to the MNIST dataset.
Specifically, PQGD with different $b$ outperforms PGD for all maximum allowed $\ell_{\infty}$ perturbations ranging 
from 0.1/1.0 to 0.2/1.0.

 We conclude that when the maximum allowed
perturbation is small, PQGD achieves comparable performance as PGD, as demonstrated in the results of Figure \ref{fig:pert} and Table \ref{tab:cifar_res}.
 In contrast when the maximum allowed perturbation increases,
PQGD creates a sharper attack than PGD. Therefore, gradient quantization is more effective when the maximum allowed 
perturbation is larger.

\subsection{Effect Of Number Of Steps On Gradient Quantization}\label{qg_step}

\begin{figure*}[h!]
\hspace{2.6cm}
\subfigure[MNIST]{\includegraphics[width=0.25\textwidth]{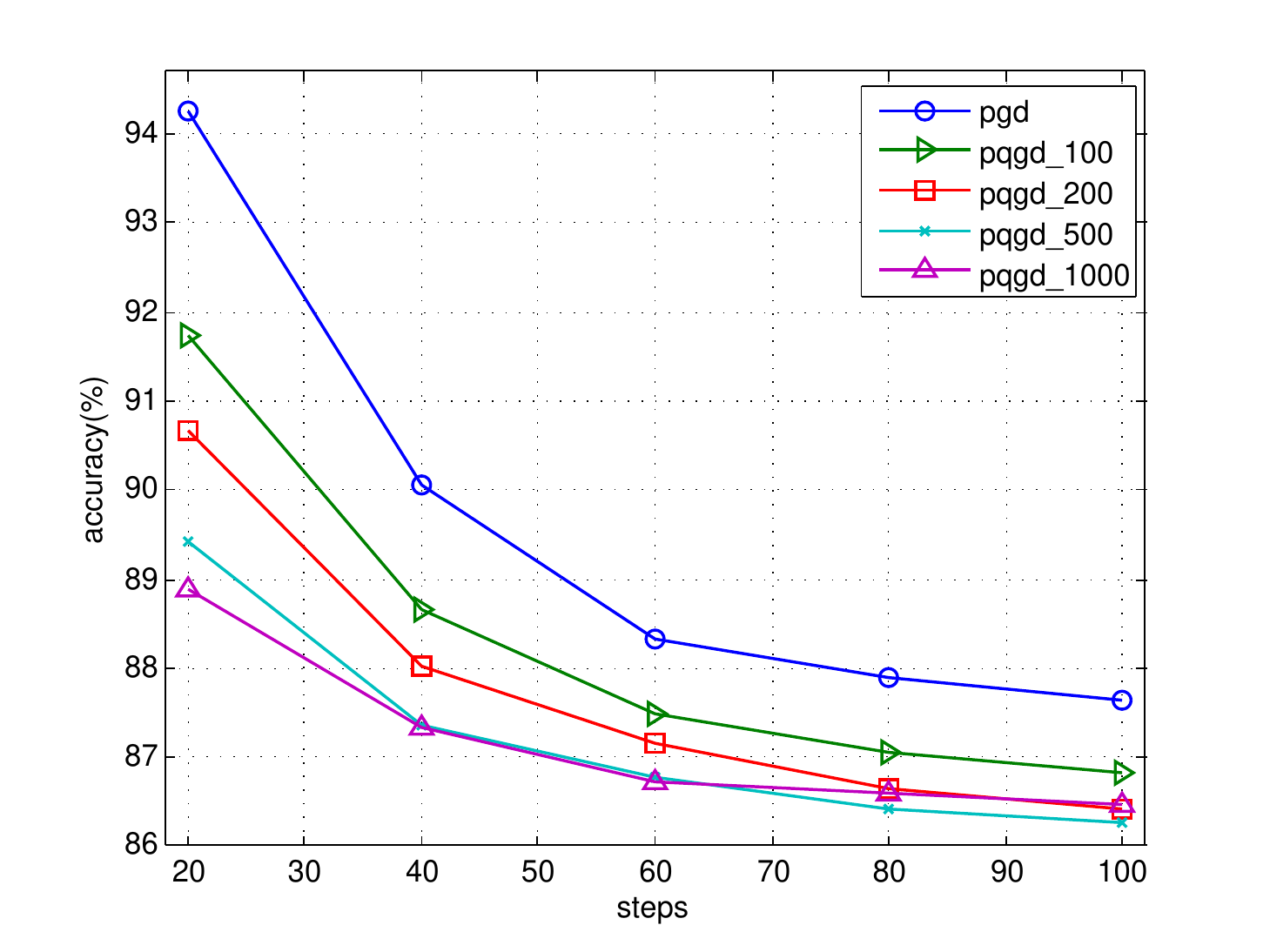} \label{fig:mnist_step}}
\hspace{3cm}
\subfigure[Fashion-MNIST]{\includegraphics[width=0.25\textwidth]{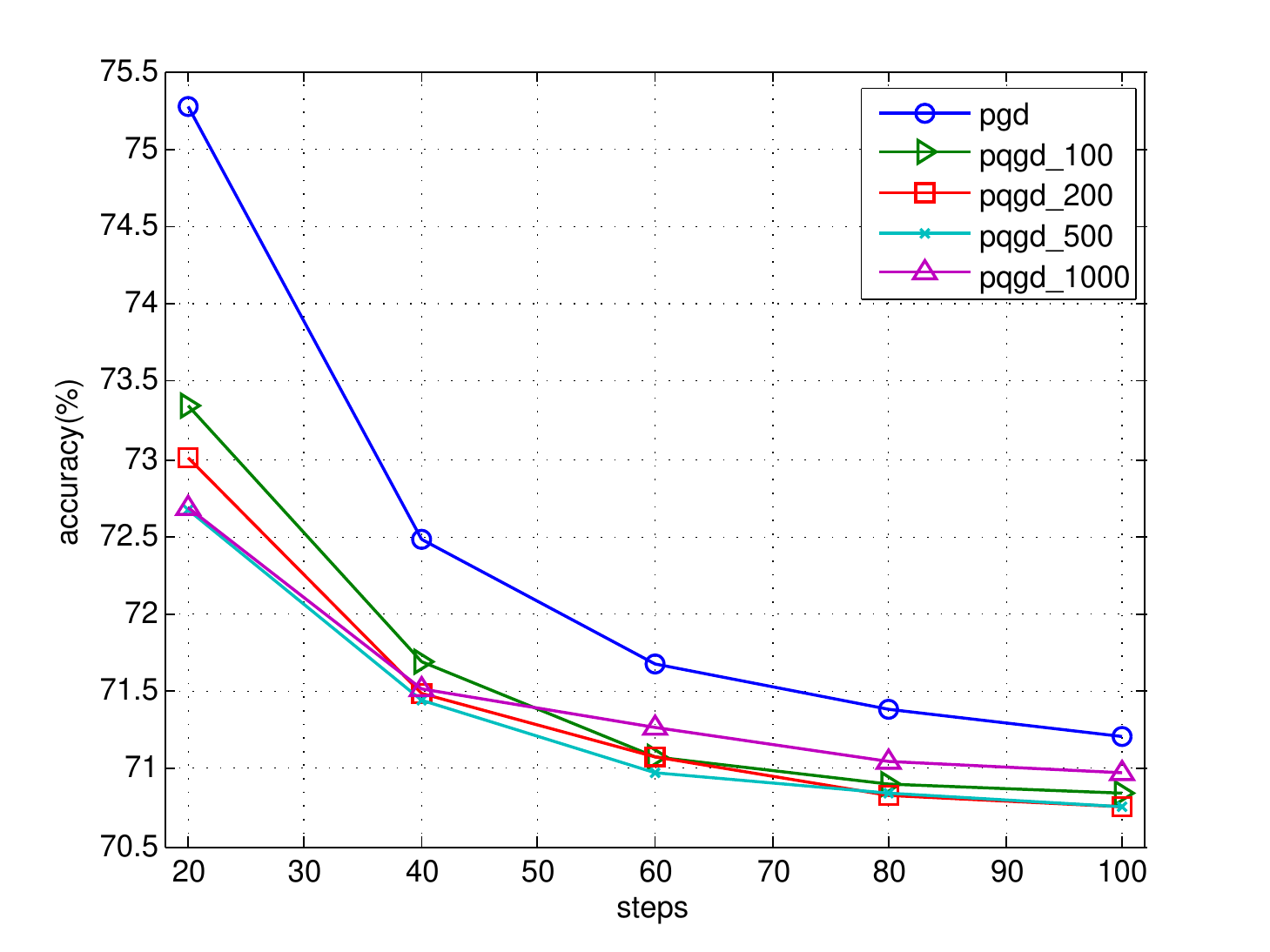} \label{fig:fmnist_step}}
\hspace{2.6cm}
\caption{Effect of the number of steps on the averaged accuracy when  applying PQGD and PGD attacks on adversarial trained models}
\label{fig:step}
\end{figure*}

We plot the average accuracy of five independent runs of PQGD and PGD, with different
steps on the MNIST and Fashion-MNIST
dataset (Fig. \ref{fig:step}). Specifically, we test PQGD and PGD under a 0.3/1.0 $\ell_{\infty}$ 
maximum allowed perturbation on the adversarial trained MNIST model while we test approaches under a 0.2/1.0  $\ell_{\infty}$ 
maximum allowed perturbation on the adversarial trained Fashion-MNIST model.

As shown in Figures \ref{fig:mnist_step} and \ref{fig:fmnist_step}, PQGD outperforms PGD with each choice of 
 $b$ and for every number of steps ranging from 20 to 100. Interestingly, PQGD produces a much sharper
 attack  than PGD when number of steps
  is small, e.g., 20 steps. It verifies our pervious claim that quantized gradients make use
of the relative magnitude to boost the efficiency of crafting adversarial examples.

In the MNIST dataset, PQGD with $b=1000$ achieves an average accuracy of $88.89\%$ 
on our adversarial trained MNIST
model while PGD reaches an average  accuracy of $94.26\%$ when number of steps is $20$.
 PGD produces similar performance results to PQGD at step $20$, when the number of steps increases to $60$.
In the Fashion-MNIST dataset, we observe similar effects of number of steps on the performance of PQGD.
When the number of steps is fixed to  $20$, PQGD with $b=500$ has an average accuracy of 
$72.68\%$ on the adversarial trained Fashion-MNIST model, while PGD has an averaged
accuracy of $75.27\%$. 

\begin{figure*}[h!]
\hspace{2.6cm}
\subfigure[8.0/255.0]{\includegraphics[width=0.25\textwidth]{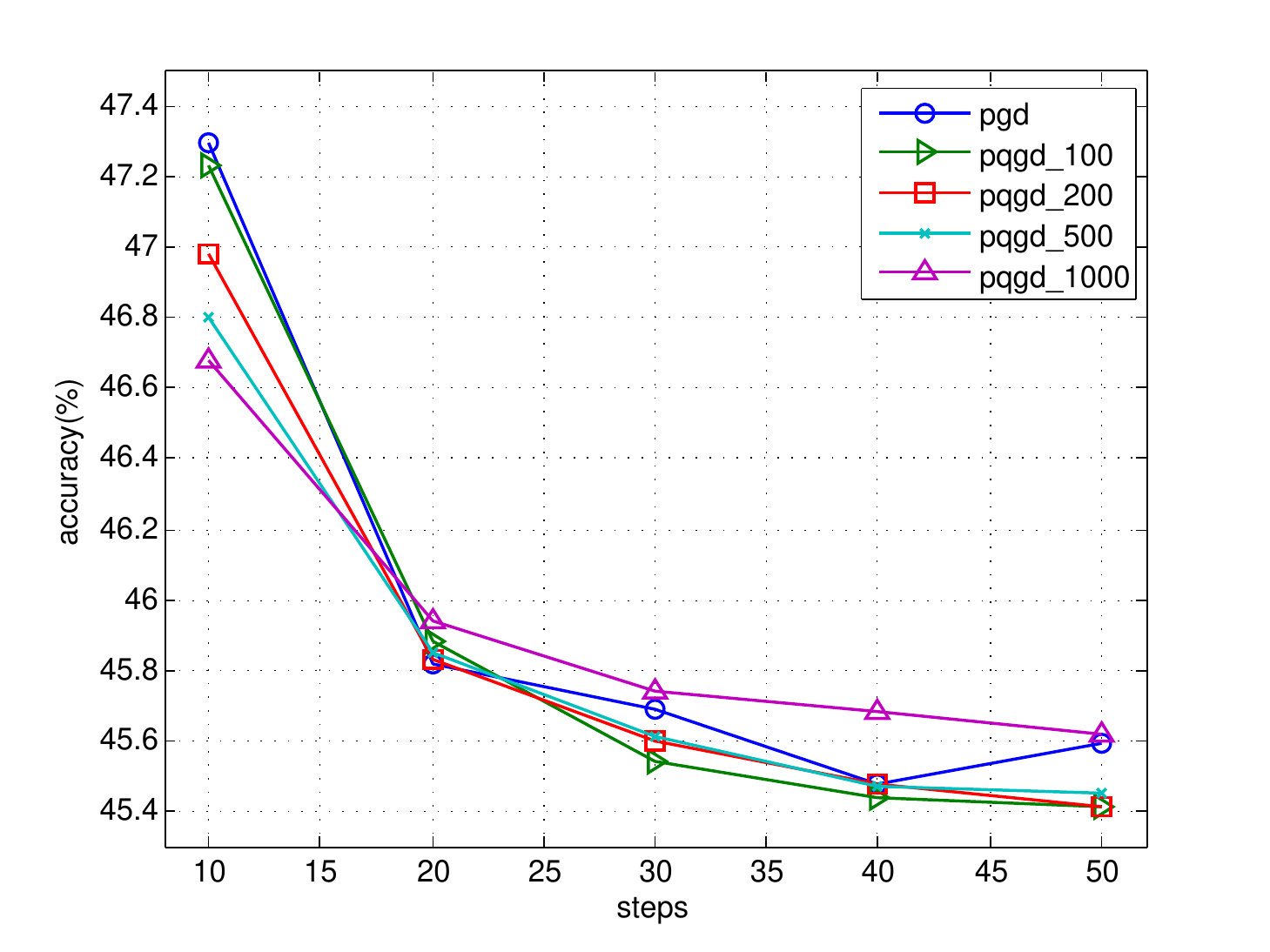} \label{fig:cifar_step_8}}
\hspace{3cm}
\subfigure[16.0/255.0]{\includegraphics[width=0.25\textwidth]{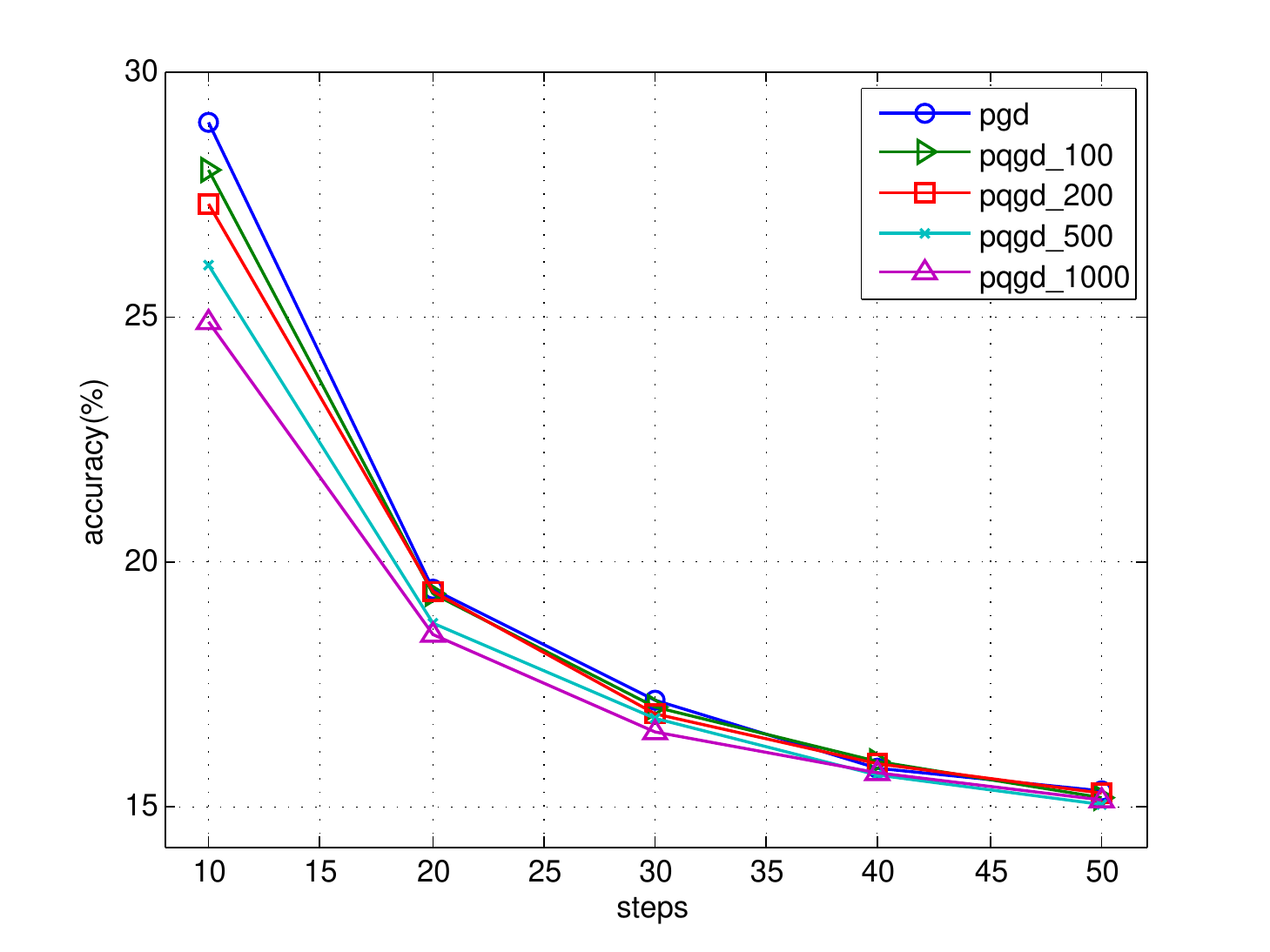} \label{fig:cifar_step_16}}
\hspace{2.6cm}
\caption{Effect of the number of steps on the accuracy of  applying PQGD and PGD attacks on adversarial trained Cifar10 models under different allowed perturbations}
\label{fig:cifar_step}
\end{figure*}

Figure \ref{fig:cifar_step} shows the effects of the number of steps on the performance of PQGD and PGD on the 
adversarial trained Cifar10 model from the Madry lab under $\ell_{\infty}$ allowed perturbations of $8.0/255.0$
and $16.0/255.0$. Under both allowed perturbations, PQGD performs better than PGD when number of steps is small,
which is consistent with the result of MNIST and Fashion-MNIST datasets.

When the  number of steps is $10$, under $8.0/255.0$ $\ell_{\infty}$ allowed perturbation, PQGD with $b=1000$ achieves
an accuracy of $46.68\%$ on the adversarial trained Cifar10 model, while PGD has an accuracy of 
$47.3\%$. On the other hand, under $16.0/255.0$ $\ell_{\infty}$ allowed perturbation, PQGD with $b=1000$
has an accuracy of $24.91\%$ on the adversarial trained Cifar10 model, while PGD has an accuracy of 
$28.98\%$. 

As the number of steps increases, the advantage of PQGD decreases compared with PGD. When the step number increases
to $40$, PQGD and PGD demonstrate a  similar performance on the adversarial trained Cifar10 model.

\subsection{Choice of Hyperparameter $b$}

The only hyperparameter we need to tune in our approach is $b$. We test PQGD and BLOB\_QG with 
a wide spectrum of $b$. In previous sections, we only present part of results because of the space
constraint. We observe that on each dataset for small $b$, i.e. $b < 100$, 
attacks with quantized gradients exhibit limited
improvement over their sign gradients. When $b$ is moderate, i.e., $ 100 \leq b \leq 1000$, 
gradient quantization achieves the best performance.  Therefore we present results when 
$b=100, 200, 500$ and  $1000$ in previous sections. When $b$ is sufficiently large,
i.e.,$b > 1000$, attacks with quantized gradients gain
advantage when the step size is small (Fig. \ref{fig:step} and \ref{fig:cifar_step}). However
when step number increases, the accuracy of attacks saturates at a relatively high level, and it  will sometimes
 performs even worse
than the vanilla PGD and DAA-BLOB approach.

\section{Related Works}

\textbf{White-box Attack:}\quad Recall the definition of the white-box attack where the attacker has complete
access to the underlying classifier information, including network architecture and parameters, training data
and labels, even the defense mechanism deployed by the underlying system.

One of the earliest white-box attacks is FGSM\cite{DBLP:journals/corr/GoodfellowSS14}, 
which is used to show  the linear
nature of DNN  causes the vulnerability of DNN to adversarial examples. While FGSM is 
an attack based on $\ell_{\infty}$ distance, 
\cite{DBLP:conf/eurosp/PapernotMJFCS16} introduced an attack under $\ell_{0}$ distance known as 
Jacobian-based Saliency Map Attack (JSMA). C\&W attack \cite{DBLP:conf/sp/Carlini017}
is introduced as an attack which can be applied 
under $\ell_{0}$, $\ell_{2}$ and $\ell_{\infty}$ distances. It is shown to significantly outperform
 FGSM and JSMA under the corresponding distance metrics. Research by \cite{DBLP:conf/iclr/TramerKPGBM18}
shows a surprising observation whereby adversarial training with single step attack such as FGSM leads
to a degenerate global minimum. They resolved the issue by using attacks crafted from different 
neural networks. PGD is the universal first-order adversary \cite{DBLP:conf/iclr/MadryMSTV18}, and a DNN
model trained with PGD is supposed to be robust to any first-order attack. 

Our work is an extension of PGD.  We replace the sign gradient in each iteration of PGD 
with quantized gradients. PQGD and BLOB\_QG belong to the category of white-box attacks because
it has access to the network parameters and architecture, training data and its corresponding label.

\textbf{Adversarial Training} \quad Adversarial training was first introduced in 
\cite{DBLP:journals/corr/GoodfellowSS14} and \cite{DBLP:conf/iclr/MadryMSTV18} who formally
define it through a  lens of modified empirical risk minimization (Eq. \ref{eq:erm}).
Evidence in \cite{DBLP:conf/icml/AthalyeC018} have shown that adversarial
 training is a state-of-the-art approach to increase 
the robustness of  a model against adversarial examples.

\textbf{Other Defenses}\quad
Gradient masking is a defense that causes the target network to generate corrupted gradients
\cite{DBLP:conf/eurosp/PapernotMJFCS16}. Existing literatures including Thermometer 
Encoding \cite{DBLP:conf/iclr/BuckmanRRG18}, Input Transformation 
\cite{DBLP:conf/iclr/GuoRCM18}, stochastic activation pruning \cite{DBLP:conf/iclr/DhillonALBKKA18}
Pre-Input Randomization Layer \cite{DBLP:conf/iclr/XieWZRY18} PixelDefend \cite{DBLP:conf/iclr/SongKNEK18},
and Defense Distillation \cite{DBLP:conf/sp/PapernotM0JS16}
fall into this category. Research by \cite{DBLP:conf/icml/AthalyeC018}
shows that there are inherent limitations of these approaches and these defenses can be bypassed
using more elaborated attack.

\section{Conclusion}

In this work, we revisit sign gradients, which are widely used in the white-box attacks, e.g., FGSM,
PGD and DAA-BLOB. We argue that existing first-order attacks with sign gradients discards the information of relative magnitude
between components in the raw gradient and thus affects the process of crafting effective
adversarial examples. We propose
quantized gradients 
 to preserve the relative magnitude between components of raw gradients
 and we integrate existing first order attacks with them. Experiments show that iterative 
first order attacks with quantized gradients outperforms the attacks with sign gradients.

{\small
\bibliographystyle{ieee_fullname}
\bibliography{egpaper_for_review}

\begin{thebibliography}{10}\itemsep=-1pt

\bibitem{DBLP:conf/nips/AlistarhG0TV17}
Dan Alistarh, Demjan Grubic, Jerry Li, Ryota Tomioka, and Milan Vojnovic.
\newblock {QSGD:} communication-efficient {SGD} via gradient quantization and
  encoding.
\newblock In {\em Advances in Neural Information Processing Systems 30: Annual
  Conference on Neural Information Processing Systems 2017, 4-9 December 2017,
  Long Beach, CA, {USA}}, pages 1707--1718, 2017.

\bibitem{DBLP:conf/icml/AthalyeC018}
Anish Athalye, Nicholas Carlini, and David~A. Wagner.
\newblock Obfuscated gradients give a false sense of security: Circumventing
  defenses to adversarial examples.
\newblock In {\em Proceedings of the 35th International Conference on Machine
  Learning, {ICML} 2018, Stockholmsm{\"{a}}ssan, Stockholm, Sweden, July 10-15,
  2018}, pages 274--283, 2018.

\bibitem{DBLP:conf/iclr/BuckmanRRG18}
Jacob Buckman, Aurko Roy, Colin Raffel, and Ian~J. Goodfellow.
\newblock Thermometer encoding: One hot way to resist adversarial examples.
\newblock In {\em 6th International Conference on Learning Representations,
  {ICLR} 2018, Vancouver, BC, Canada, April 30 - May 3, 2018, Conference Track
  Proceedings}, 2018.

\bibitem{DBLP:conf/sp/Carlini017}
Nicholas Carlini and David~A. Wagner.
\newblock Towards evaluating the robustness of neural networks.
\newblock In {\em 2017 {IEEE} Symposium on Security and Privacy, {SP} 2017, San
  Jose, CA, USA, May 22-26, 2017}, pages 39--57, 2017.

\bibitem{DBLP:conf/iclr/DhillonALBKKA18}
Guneet~S. Dhillon, Kamyar Azizzadenesheli, Zachary~C. Lipton, Jeremy Bernstein,
  Jean Kossaifi, Aran Khanna, and Animashree Anandkumar.
\newblock Stochastic activation pruning for robust adversarial defense.
\newblock In {\em 6th International Conference on Learning Representations,
  {ICLR} 2018, Vancouver, BC, Canada, April 30 - May 3, 2018, Conference Track
  Proceedings}, 2018.

\bibitem{DBLP:journals/corr/EvtimovEFKLPRS17}
Ivan Evtimov, Kevin Eykholt, Earlence Fernandes, Tadayoshi Kohno, Bo Li, Atul
  Prakash, Amir Rahmati, and Dawn Song.
\newblock Robust physical-world attacks on machine learning models.
\newblock {\em CoRR}, abs/1707.08945, 2017.

\bibitem{DBLP:journals/corr/GoodfellowSS14}
Ian~J. Goodfellow, Jonathon Shlens, and Christian Szegedy.
\newblock Explaining and harnessing adversarial examples.
\newblock In {\em 3rd International Conference on Learning Representations,
  {ICLR} 2015, San Diego, CA, USA, May 7-9, 2015, Conference Track
  Proceedings}, 2015.

\bibitem{gowal2019alternative}
Sven Gowal, Jonathan Uesato, Chongli Qin, Po-Sen Huang, Timothy Mann, and
  Pushmeet Kohli.
\newblock An alternative surrogate loss for pgd-based adversarial testing.
\newblock {\em arXiv preprint arXiv:1910.09338}, 2019.

\bibitem{DBLP:conf/iclr/GuoRCM18}
Chuan Guo, Mayank Rana, Moustapha Ciss{\'{e}}, and Laurens van~der Maaten.
\newblock Countering adversarial images using input transformations.
\newblock In {\em 6th International Conference on Learning Representations,
  {ICLR} 2018, Vancouver, BC, Canada, April 30 - May 3, 2018, Conference Track
  Proceedings}, 2018.

\bibitem{krizhevsky2009learning}
Alex Krizhevsky and Geoffrey Hinton.
\newblock Learning multiple layers of features from tiny images.
\newblock Technical report, Citeseer, 2009.

\bibitem{DBLP:conf/iclr/KurakinGB17}
Alexey Kurakin, Ian~J. Goodfellow, and Samy Bengio.
\newblock Adversarial machine learning at scale.
\newblock In {\em 5th International Conference on Learning Representations,
  {ICLR} 2017, Toulon, France, April 24-26, 2017, Conference Track
  Proceedings}, 2017.

\bibitem{lecun1998gradient}
Yann LeCun, L{\'e}on Bottou, Yoshua Bengio, Patrick Haffner, et~al.
\newblock Gradient-based learning applied to document recognition.
\newblock {\em Proceedings of the IEEE}, 86(11):2278--2324, 1998.

\bibitem{DBLP:conf/iclr/MadryMSTV18}
Aleksander Madry, Aleksandar Makelov, Ludwig Schmidt, Dimitris Tsipras, and
  Adrian Vladu.
\newblock Towards deep learning models resistant to adversarial attacks.
\newblock In {\em 6th International Conference on Learning Representations,
  {ICLR} 2018, Vancouver, BC, Canada, April 30 - May 3, 2018, Conference Track
  Proceedings}, 2018.

\bibitem{DBLP:journals/corr/PapernotMG16}
Nicolas Papernot, Patrick~D. McDaniel, and Ian~J. Goodfellow.
\newblock Transferability in machine learning: from phenomena to black-box
  attacks using adversarial samples.
\newblock {\em CoRR}, abs/1605.07277, 2016.

\bibitem{DBLP:conf/ccs/PapernotMGJCS17}
Nicolas Papernot, Patrick~D. McDaniel, Ian~J. Goodfellow, Somesh Jha, Z.~Berkay
  Celik, and Ananthram Swami.
\newblock Practical black-box attacks against machine learning.
\newblock In {\em Proceedings of the 2017 {ACM} on Asia Conference on Computer
  and Communications Security, AsiaCCS 2017, Abu Dhabi, United Arab Emirates,
  April 2-6, 2017}, pages 506--519, 2017.

\bibitem{DBLP:conf/eurosp/PapernotMJFCS16}
Nicolas Papernot, Patrick~D. McDaniel, Somesh Jha, Matt Fredrikson, Z.~Berkay
  Celik, and Ananthram Swami.
\newblock The limitations of deep learning in adversarial settings.
\newblock In {\em {IEEE} European Symposium on Security and Privacy, EuroS{\&}P
  2016, Saarbr{\"{u}}cken, Germany, March 21-24, 2016}, pages 372--387, 2016.

\bibitem{DBLP:conf/sp/PapernotM0JS16}
Nicolas Papernot, Patrick~D. McDaniel, Xi Wu, Somesh Jha, and Ananthram Swami.
\newblock Distillation as a defense to adversarial perturbations against deep
  neural networks.
\newblock In {\em {IEEE} Symposium on Security and Privacy, {SP} 2016, San
  Jose, CA, USA, May 22-26, 2016}, pages 582--597, 2016.

\bibitem{DBLP:conf/ccs/SharifBBR16}
Mahmood Sharif, Sruti Bhagavatula, Lujo Bauer, and Michael~K. Reiter.
\newblock Accessorize to a crime: Real and stealthy attacks on state-of-the-art
  face recognition.
\newblock In {\em Proceedings of the 2016 {ACM} {SIGSAC} Conference on Computer
  and Communications Security, Vienna, Austria, October 24-28, 2016}, pages
  1528--1540, 2016.

\bibitem{DBLP:conf/iclr/SongKNEK18}
Yang Song, Taesup Kim, Sebastian Nowozin, Stefano Ermon, and Nate Kushman.
\newblock Pixeldefend: Leveraging generative models to understand and defend
  against adversarial examples.
\newblock In {\em 6th International Conference on Learning Representations,
  {ICLR} 2018, Vancouver, BC, Canada, April 30 - May 3, 2018, Conference Track
  Proceedings}, 2018.

\bibitem{DBLP:journals/corr/SzegedyZSBEGF13}
Christian Szegedy, Wojciech Zaremba, Ilya Sutskever, Joan Bruna, Dumitru Erhan,
  Ian~J. Goodfellow, and Rob Fergus.
\newblock Intriguing properties of neural networks.
\newblock In {\em 2nd International Conference on Learning Representations,
  {ICLR} 2014, Banff, AB, Canada, April 14-16, 2014, Conference Track
  Proceedings}, 2014.

\bibitem{DBLP:conf/iclr/TramerKPGBM18}
Florian Tram{\`{e}}r, Alexey Kurakin, Nicolas Papernot, Ian~J. Goodfellow, Dan
  Boneh, and Patrick~D. McDaniel.
\newblock Ensemble adversarial training: Attacks and defenses.
\newblock In {\em 6th International Conference on Learning Representations,
  {ICLR} 2018, Vancouver, BC, Canada, April 30 - May 3, 2018, Conference Track
  Proceedings}, 2018.

\bibitem{DBLP:journals/corr/abs-1811-02625}
Shiqi Wang, Yizheng Chen, Ahmed Abdou, and Suman Jana.
\newblock Mixtrain: Scalable training of formally robust neural networks.
\newblock {\em CoRR}, abs/1811.02625, 2018.

\bibitem{xiao2017fashion}
Han Xiao, Kashif Rasul, and Roland Vollgraf.
\newblock Fashion-mnist: a novel image dataset for benchmarking machine
  learning algorithms.
\newblock {\em arXiv preprint arXiv:1708.07747}, 2017.

\bibitem{DBLP:conf/iclr/XieWZRY18}
Cihang Xie, Jianyu Wang, Zhishuai Zhang, Zhou Ren, and Alan~L. Yuille.
\newblock Mitigating adversarial effects through randomization.
\newblock In {\em 6th International Conference on Learning Representations,
  {ICLR} 2018, Vancouver, BC, Canada, April 30 - May 3, 2018, Conference Track
  Proceedings}, 2018.

\bibitem{DBLP:journals/corr/abs-1808-05537}
Tianhang Zheng, Changyou Chen, and Kui Ren.
\newblock Distributionally adversarial attack.
\newblock {\em CoRR}, abs/1808.05537, 2018.

\end{thebibliography}
}

\end{document}